# Gate-tunable Veselago Interference in a Bipolar Graphene Microcavity

Xi Zhang[1†], Wei Ren[1†], Elliot Bell[1], Ziyan Zhu[2], Kenji Watanabe[3], Takashi Taniguchi[4], Efthimios Kaxiras[2,5], Mitchell Luskin[6], Ke Wang[1*]

[1]*School of Physics and Astronomy, University of Minnesota, Minneapolis, MN, USA*

[2]*Department of Physics, Harvard University, Cambridge, MA, USA*

[3]*Research Center for Functional Materials, National Institute for Materials Science, Tsukuba, Ibaraki, Japan*

[4]*International Center for Materials Nanoarchitectonics, National Institute for Materials Science, Tsukuba, Ibaraki, Japan*

[5]*John A. Paulson School of Engineering and Applied Sciences, Harvard University, Cambridge, Massachusetts, USA*

[6]*School of Mathematics, University of Minnesota, Minneapolis, MN, USA*

**The relativistic charge carriers in monolayer graphene can be manipulated in manners akin to conventional optics[1,2] (electron-optics): angle-dependent Klein tunneling[3,4] collimates an electron beam (analogous to a laser), while a Veselago refraction process focuses it (analogous to an optical lens). Both processes have been previously investigated[5–10], but the collimation and focusing efficiency have been reported to be relatively low even in state-of-the-art ballistic pn-junction devices. These limitations prevented the realization of more advanced quantum devices based on electron-optical interference, while understanding of the underlying physics remains elusive. Here, we present a novel device architecture of a graphene microcavity defined by carefully-engineered local strain and electrostatic fields. We create a controlled electron-optic interference process at zero magnetic field as a consequence of consecutive Veselago refractions in the microcavity and provide direct experimental evidence through low-temperature electrical transport measurements. The experimentally observed first-, second-, and third-order interference peaks agree quantitatively with the Veselago physics in a microcavity. In addition, we demonstrate decoherence of the interference by an external magnetic field, as the cyclotron radius becomes comparable to the interference length scale. For its application in electron-optics, we utilize Veselago interference to localize uncollimated electrons and characterize its**

*Corresponding author: kewang@umn.edu
† These authors contributed equally

**contribution in further improving collimation efficiency. Our work sheds new light on relativistic single-particle physics and provides important technical improvements toward next-generation quantum devices based on the coherent manipulation of electron momentum and trajectory.**

The linear dispersion relationship of electrons in monolayer graphene[11,12] permits the manipulation of ballistic electron trajectories[13–16] in a manner akin to classical optics. It has been demonstrated that an electron passing through a graphene pn-junction[17] can be collimated[18,19] through an angle-dependent Klein tunneling process[7] (analogous to a laser) and can also be refracted by a Veselago lensing process[20,21] (analogous to an optical lens) depending on the width and height of the pn-junction barrier. However, in state-of-the-art graphene pn-junctions, there still exist important technical challenges for creating a full electronic version of advanced optical circuits. First, due to the relatively small pseudo-gap that can be created by an electrostatically-defined pn-junction, the Klein tunneling probability across the junction does not depend sensitively on the incident angle. In addition to electrons with perpendicular momentum to the junction that flow across it effortlessly, electrons with finite incident angle $\theta$ relative to the perpendicular axis have a non-trivial transmission probability, limiting the resulting collimation efficiency. Moreover, the flatness of the charge neutrality boundary at which the charge carrier type switches in the pn-junction is highly sensitive to charge inhomogeneity (even in the highest quality devices), introducing undesirable astigmatism analogous to a deformed lens. These limitations present significant challenges in studying more complex electron-optics processes such as controlled quantum interference, and in developing future quantum electronic devices based on more advanced electron-optical circuits.

In this work, we present a novel device architecture of a bipolar graphene microcavity that addresses these challenges with precise strain and electrostatic engineering. In this new device platform, we demonstrate novel electro-optics phenomena resulting from interference through consecutive Veselago lensing processes[22,23], referred to as "Veselago interference." We study the electrostatic and magnetic field dependence using low-temperature transport measurements and demonstrate quantitative agreement to relativistic single-particle physics. Finally, by utilizing the Veselago interference demonstrated here, we further localize uncollimated electrons to improve collimation efficiency, providing proof-of-concept demonstration of a new collimation scheme aimed toward future electro-optical devices.

A piece of monolayer graphene (MLG) is encapsulated by two layers of hexagonal boron nitride (hBN) using the standard dry-transfer technique[24–26]. The stack is subsequently transferred on top of two pre-patterned local bottom gates with a ~50 nm lateral gap, and an intentional 8 nm vertical height difference (fig. 1a). Fig. 1b shows an optical microscope image of the complete device. Atomic-force-microscopy data (fig. 1c) across the 50 nm gapped region reveals that the device remains atomically flat on top of both gates as well as in the gap, while lattice distortions are apparent at the boundaries of both gates (see SI for more information about device fabrication). The strain-induced band gap[27,28] (fig. 1d) effectively defines two tunnel barriers.

For large (zero) incident angles, each strain-induced barrier provides near-perfect reflection (transmission) as expected from Klein tunneling angle dependence. For carriers with small but finite incident angle $\theta$, these two tunnel barriers effectively define a microcavity in between, where the injection rate to and loss rate from the cavity are equally low. The confinement of carriers with small incident angles is a meticulous experimental design to isolate, study, and manipulate those carriers of interest, allowing an interference loop to form via two consecutive Veselago refractions inside the cavity. In terms of application, these carriers are primarily responsible for the relatively low collimation efficiency previously reported. Further suppression of their contribution to electrical transport could be the key to improve collimation efficiency for future electron-optics devices.

The two local bottom gates, when biased at opposite voltages (of equal magnitude), electrostatically define a pn-junction symmetric against the charge neutrality boundary in the center of the cavity. The resulting carrier density distribution and the trajectory of carriers inside the cavity are shown in the color diagram of fig. 1a. After being injected into the cavity with a small incident angle (and probability $p \ll 1$), each carrier inside the cavity first undergoes Veselago refraction as the carrier type changes in the middle of the device, gets reflected from the cavity boundary on the opposite side (with probability $1 - p \sim 1$), then undergoes a second Veselago refraction that ultimately brings the carrier back to its original position. At zero magnetic field, the charge constructively interferes with itself. This results in an increase in measured resistance similar to weak localization (fig. 1e), but with two major differences. First, the effect is not weak, as all electrons trapped in the cavity form constructive interference loops, independent of the specific incident angle. Thus the cavity is also efficient in localizing all uncollimated electrons.

Second, the effect can be turned onto or off resonance by tuning inference paths with electrostatics, without the need for a magnetic field to suppress weak localization.

Fig. 1e shows the measured resistance as a function of the densities of the two adjacent bottom gate regions, with $n_1$ and $n_2$ corresponding to the carrier density in the Gate 1 and Gate 2 regions, respectively. Resistance peak traces are observed only when the carrier densities on top of Gates 1 and 2 are of opposite charge (when a pn-junction is formed within the cavity). The absence of these transport features in the nn or pp gate configurations rule out the possibility for them to arise from disorder or Fabry-Pérot interference. The ballistic (or disorder-free) nature of the electron is also supported by the high electron mobility (of ~ 300,000 cm$^2$/Vs).

Figure 2a-c shows the measured 4-probe resistance of three different microcavities in the regime as a function of the carrier density above Gate 1 and Gate 2. Resistance peaks are visible at $|n_1| = |n_2|$ (red diamond), $|n_1| = 4|n_2|$ (green square), $4|n_1| = |n_2|$ (blue circle). At $n_1/n_2 = 1/l^2$ or $l^2$ for integers $l$, a charge carrier can undergo two Veselago refractions and $l$-1 reflections at the charge neutrality boundary (fig. 2 d-h), forming a closed-loop trajectory back to its original position (see SI for detailed analysis), constructively interfering with itself at the first cavity wall where it was originally injected. When the charge eventually leaves the cavity, the Veselago interference results in a higher chance of the electron escaping via the first cavity wall (reflected back to the source, and does not contribute to the current to the drain) than the opposite cavity wall (toward the drain, contributing to an uncollimated current to drain). In fig. 2d, the bold arrow marks the main inference loop of the first-order interference peak as a consequence of two consecutive Veselago refractions. The faded line marks the trajectory of a smaller portion of carriers that are reflected (instead of refracted) at the pn-junction boundary in the center of the cavity, eventually contributing to the same resistance peak via two consecutive Veselago refractions, to the side of the main loop (see SI for more information). Similarly, a closed interference loop (bold) can be formed via a sequence of two refractions and one reflection at $|n_1| = 4|n_2|$ and $4|n_1| = |n_2|$, leading to second-order Veselago interference peaks (see SI for more information). The detailed features of the resonance peak are also consistent with Veselago physics under the influence of realistic device details such as the finite width and inhomogeneity of the strain-defined barrier (see SI for more information).

To further confirm the nature of Veselago interference, we measure the 4-probe resistance of the cavity under an out-of-plane external magnetic field. At small magnetic field of $B$ = 0.3 T, the Lorentz force prevents carriers injected at relatively large angle from reaching the charge-neutrality boundary and disrupts the interference loop for carriers that do manage to reach the boundary at smaller incident angle (fig. 3a). This results in a broadening of the resistance peaks, as the new adjusted interference condition differs by specific incident angle (fig. 3b). When the cyclotron orbit radius becomes comparable to the width of the cavity at $B = 0.5$ T, no injected carriers are expected to reach the charge-neutrality boundary, and no Veselago refraction occurs (see SI for more details). As a result, the Veselago interference peaks are fully suppressed (fig. 3c). Figure 3d shows the measured resistance as a function of magnetic field and carrier density around the peak position, where $\Delta n = 0$ corresponds to the peak position. The width of the interference peak monotonically increases with magnetic field until it completely disappears. Due to the symmetry of the interference loop at $|n_1|=|n_2|$, no bias dependence is observed for the first-order peak (fig. 3e). No significant temperature dependence is observed from $T = 4$ K ~ 12 K, as the specifics of the interference loop are expected to be insensitive to thermal excitations (fig. 3f) (see SI for more details).

The angle dependence of the Klein tunneling probability has been utilized to collimate electron flow in ballistic graphene devices[3,7]. However, carriers with small-incident angle can still pass through a series of Klein barriers with a non-negligible probability. This has been a major challenge for further improving collimation efficiency for advanced electron optics. Here, we show that Veselago interference can be used to further collimate carriers, particularly those with small incident angle. Figure 4a shows the device diagram for characterizing the collimation efficiency of Veselago interference. Source (drain) contacts are intentionally placed at the bottom right (left) corners of graphene on top of Gate 2 (1). Two 1μm-wide voltage probes are placed at the top and bottom edges of the graphene on top of Gate 1. Carriers are injected from the source contact and collimated by the cavity wall, allowing only a small portion of carriers with small incident angle to enter the cavity. When $|n_1| < |n_2|$, the uncollimated carriers that escape the cavity predominantly reach voltage probe B, resulting in a transverse voltage proportional to the current density from uncollimated charge carriers (fig. 4b). At $4|n_1| = |n_2|$, Veselago interference further localizes uncollimated carriers in the cavity (fig. 4b), resulting in a near-zero measured transverse voltage amongst a high resistance background off-resonance, demonstrating its efficiency in further collimating small-incident-angle carriers that would otherwise manage to get across the cavity. For the other two interference peaks at $|n_1| = 4|n_2|$ and $|n_1| = |n_2|$, the collimation efficiency is not characterized by the transverse voltage (fig. 4c), as the ballistic carriers (collimated or not) will reach and diffusively scatter from the physical edges of the device before reaching either voltage probe (see SI for more details).

A weak bias dependence (fig. 4e) is observed for the zero transverse-resistance dip, as expected from the bias dependence of second-order peaks at $|n_1| = 4|n_2|$. As a perpendicular magnetic field is applied beyond $B = 0.3$ T, the transverse voltage increases both due to the destruction of Veselago interference as well as the curved trajectory of carriers after passing through the cavity (fig. 4f). No significant temperature dependence is recorded from $T = 4$ K ~ 12 K (fig. 4g), also consistent with our previous observations.

In conclusion, we have developed a novel device architecture for a microcavity in monolayer graphene with strain and electrostatic-field engineering. We report a new electro-optics phenomenon: Veselago interference as a result of electron localization after two consecutive Veselago refractions in the cavity. The observed low-temperature magneto-transport signature agrees quantitatively with Veselago physics and provides further experimental evidence and insights into relativistic single-particle dynamics. Finally, the electrons that participate in Veselago interference in the cavity are those with small-incident-angle momenta, the exact same electrons primarily responsible for the relatively low collimation efficiency in previously studied pn-junction Klein collimators. By characterizing the collimation efficiency with transverse voltage probes, we provide proof-of-principle demonstration that Veselago interference helps to further localize these uncollimated electrons and thus improves the collimation of current through the cavity. Our work provides an important new technical and conceptual component toward advanced electron-optic circuits based on the precise manipulation of ballistic electron trajectories.

**Data availability**

Source data for figures (including supplementary figures) are available from the corresponding author upon reasonable request.

**Code availability**

All relevant code needed to evaluate the conclusions in the paper are available from the corresponding author upon reasonable request.

**Acknowledgements**

We thank Boris Shklovskii, Alex Kamenev, and Kan-Ting Tsai for useful discussions and Atomic-Force Microscope Analysis. The work at UMN was supported by the National Science Foundation CAREER Award NSF-1944498. Portions of the UMN work were conducted in the Minnesota Nano Center, which is supported by the National Science Foundation through the National Nano Coordinated Infrastructure Network (NNCI) under Award Number ECCS-1542202. The bandstructure calculation by Z.Z., M. L. and E. K. is supported by STC Center for Integrated Quantum Materials, NSF Grant No. DMR-1231319, ARO MURI Grant No. W911NF14-0247, and NSF DMREF Grant No. 1922165. Calculations were performed on the Odyssey cluster supported by the FAS Division of Science, Research Computing Group at Harvard University and the National Energy Research Scientific Computing Center (NERSC), a U.S. Department of Energy Office of Science User Facility located at Lawrence Berkeley National Laboratory, operated under Contract No. DE-AC02-05CH11231. Portions of the hexagonal boron nitride material used in this work were provided by K.W and T. T. K.W. and T.T. acknowledge support from the Elemental Strategy Initiative conducted by the MEXT, Japan, Grant Number JPMXP0112101001 and JSPS KAKENHI Grant Numbers JP20H00354.

**Author contributions**

K.W. designed the experiment and device architecture. Data presented in this work was taken by X.Z. and W.R. Device fabrication, measurement and data analysis was performed by X.Z. and W.R. with support from E.B. under the supervision of K.W. The band structure calculation for strained graphene was performed by Z.Z. under the supervision of E.K. and M.L. A portion of the hexagonal boron nitride used in this work was provided by T.T. and K.W. X.Z., W.R., E.B., and K.W. wrote the paper. All authors discussed the results and provided comments on the manuscript.

**Competing interests**

The authors declare no competing interests.

## Additional information

**Supplementary information** is available for this paper at [URL inserted by publisher]

**Correspondence and requests for materials** should be addressed to K.W. (kewang@umn.edu)

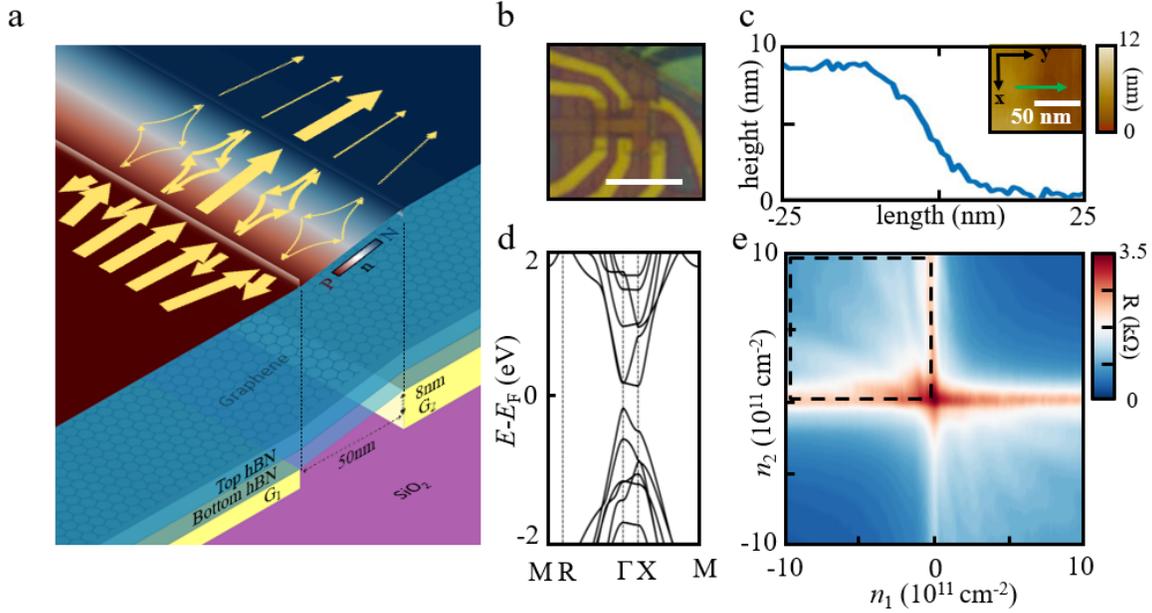

**Figure 1 | Strain-defined and Gate-tunable Veselago Quantum Interferometer.** (a) (bottom) A hBN-encapsulated monolayer graphene stack is transferred on top of two atomically-flat local bottom gates with height difference of 8 nm and lateral separation of 50 nm. (Top) The color plot maps the carrier density distribution in the device. Uniform P (N) type doping is found in regions on top of Gate 1 (2), while a uniform pn-junction forms in the cavity between the two gates. Two narrow depletion regions form along the boundaries of the gates, where significant lattice distortion opens a band gap. (Arrows) When the density of hole (electron) type carriers on top of Gate 1 (2) becomes equal, electrons entering the cavity are localized via two consecutive Veselago refractions back to their original position. (b) Optical image of a typical device. (c) AFM topology across the cavity. The 8 nm height difference and stack distortion near the boundaries are clearly visible. (d) Band structure of locally-strained monolayer graphene along the high symmetry line (see Supplemental Materials for details) shows a band gaps ~0.1 eV. (e) The measured resistance as a function of the electron density on Gate 1 ($n_1$) and Gate 2 ($n_2$), where negative electron density corresponds to hole-type doping. Resistance peaks are observed exclusively when $n_1$ and $n_2$ have opposite signs.

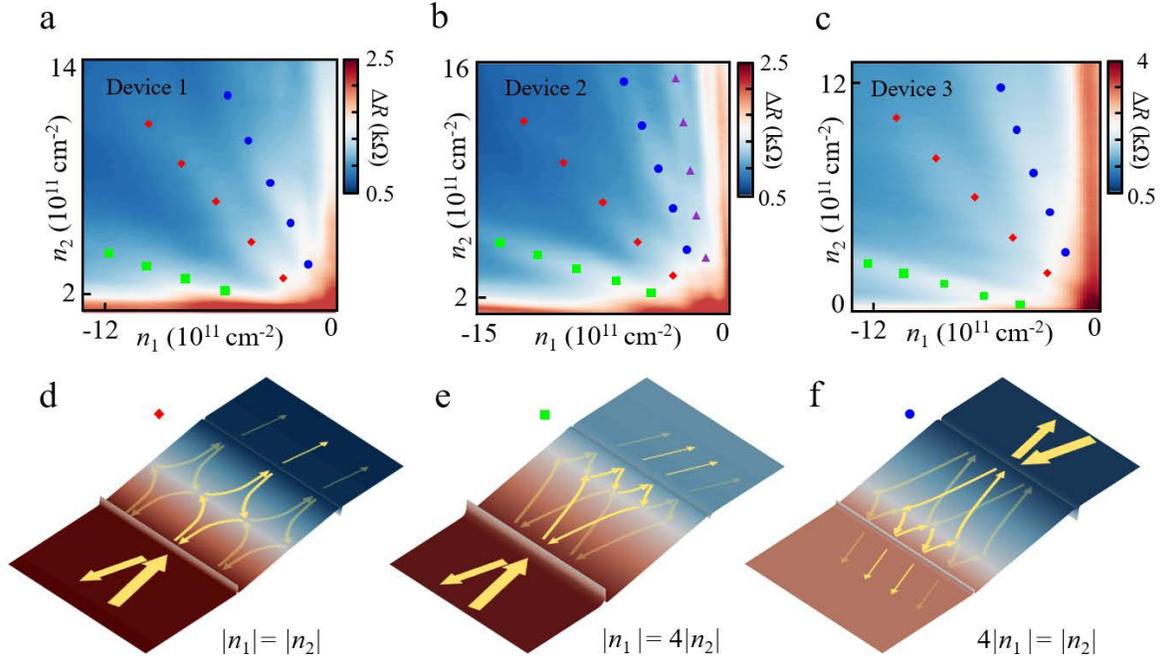

**Figure 2 | Veselago Interference and Localization.** (a-c) Measured 4-probe resistance of cavity as a function of $n_1$ and $n_2$, in three devices. Resistance peaks are visible at $|n_1| = |n_2|$ (red), $|n_1| = 4|n_2|$ (green), $4|n_1| = |n_2|$ (blue). In Device 2, a third-order resistance peak at $9|n_1| = |n_2|$ is visible. (d) When $|n_1| = |n_2|$, a symmetric pn-junction is defined in the cavity. Carriers injected into the cavity (with low probability $p$ and small incident angle) undergo Veselago refraction as the carrier type changes in the middle of the device, are reflected from the cavity boundary on the opposite side (with probability $1 - p \sim 1$), and undergo a second Veselago refraction to ultimately bring them back to their original position. At $B=0$, this constructive interference localizes the carrier, increasing the probability of the carrier leaving the cavity via the injection barrier, thus increasing the measured resistance. The bold arrow marks the main inference loop (two consecutive Veselago refractions). The faded lines mark the trajectory of a smaller portion of carriers reflected (not refracted) at the pn-junction boundary in the center of the cavity, eventually contributing to the same resistance peak via two consecutive Veselago refraction, to the side of the main loop. (e)(f) Similarly, a closed interference loop (bold) can be formed via a sequence of two refractions and one reflection at $|n_1| = 4|n_2|$ and $4|n_1| = |n_2|$, leading to a second-order Veselago interference peak.

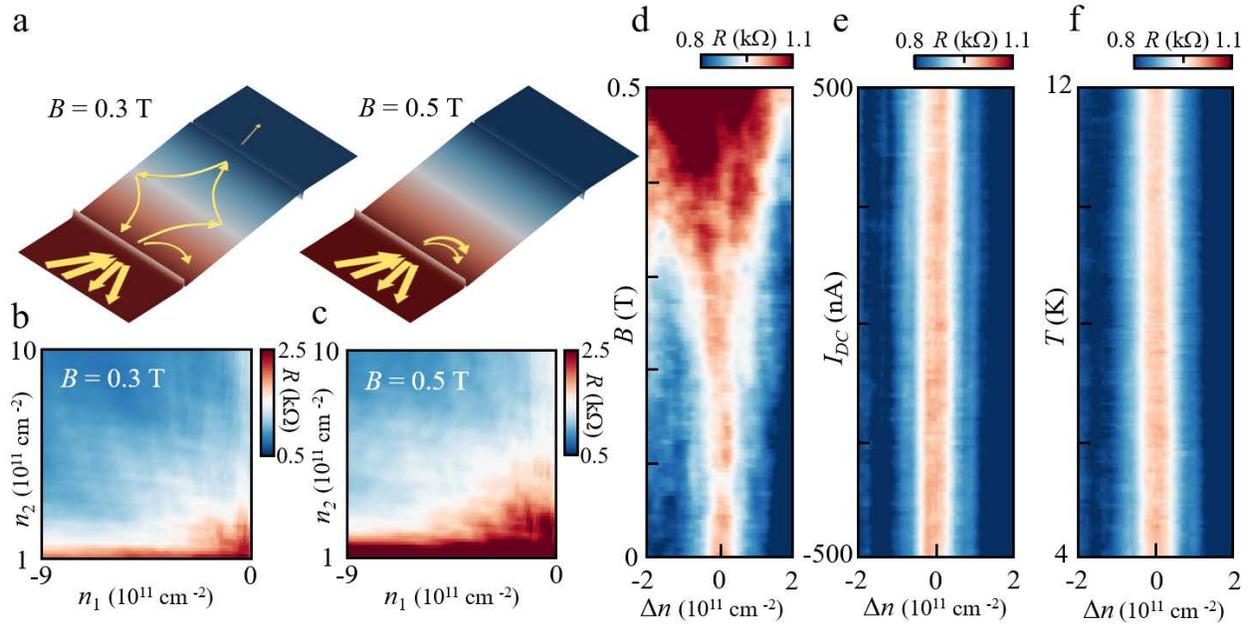

**Figure 3 | Dependence on Magnetic Field, Bias and Temperature.** (a) At $B = 0.3$ T, the Lorentz force prevents carriers injected at relatively large angles from reaching the charge neutrality boundary while shifting the interference loop for carriers managing to reach it (at smaller incident angle). (b) This results in a broadening of the resistance peaks as the new adjusted interference condition differs by specific incident angle. (c) When the cyclotron orbit radius becomes comparable to the width of the cavity at $B = 0.5$ T, no injected carriers is expected to reach the charge-neutral boundary, and therefore no Veselago refraction can occur. As a result, the Veselago interference peaks are fully suppressed. (d) Measured resistance as a function of the magnetic field and carrier density around the peak position. The width of the interference peak monotonically increases with magnetic field, until it completely disappears. (e) No bias dependence is observed for the first-order peak at $|n_1| = |n_2|$ due to the symmetry of the interference loop. (f) No significant temperature dependence is observed from $T = 4$ K ~ 12 K, as the interference loop is expected to be insensitive to thermal excitations.

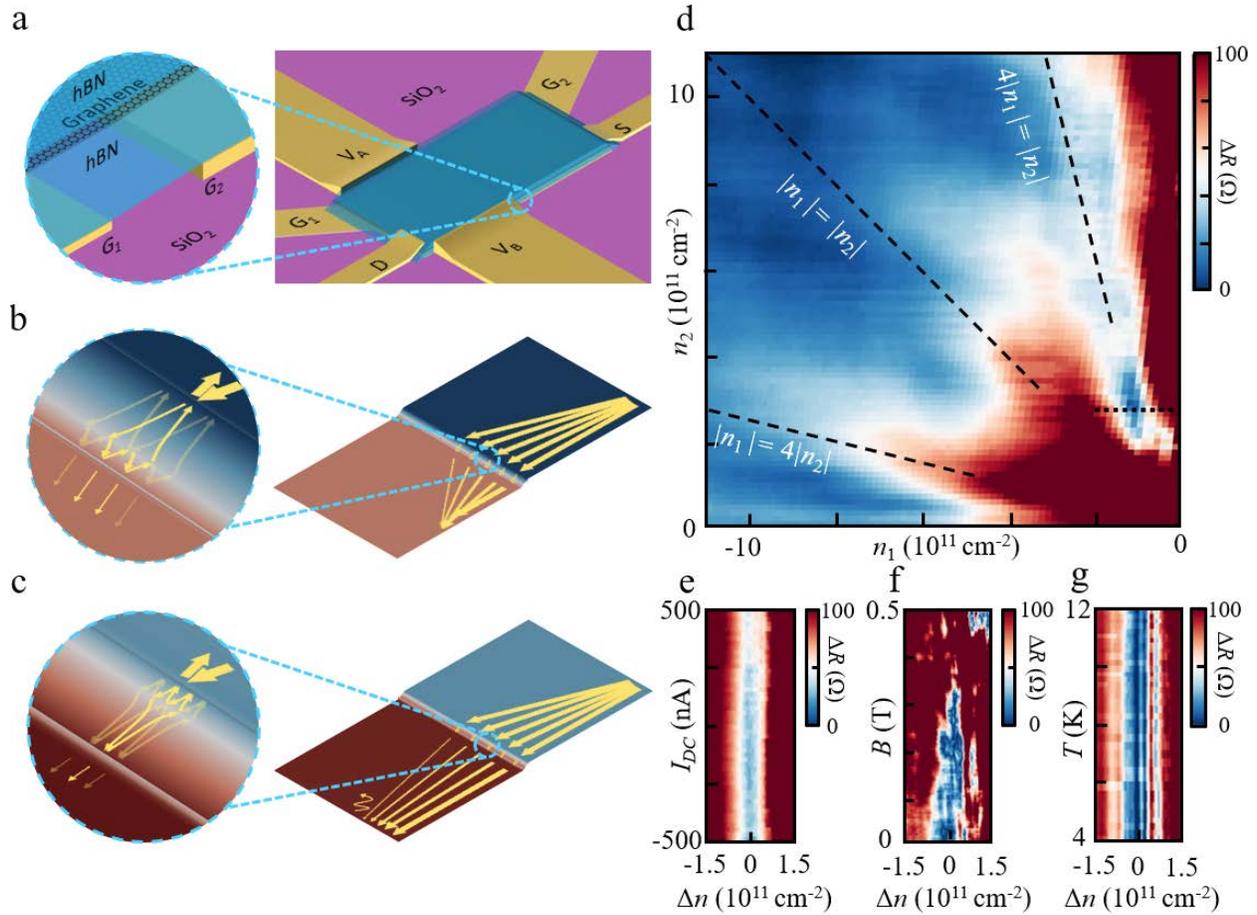

**Figure 4 | Enhancing Collimation with second-order Veselago Interference.** (a) Device diagram demonstrating the collimation of small-incident-angle electrons with Veselago interference. Source (drain) contacts are intentionally placed at the bottom right (left) corners of the graphene on top of Gate 2 (1). Two 1 μm-wide voltage probes are placed at the top and bottom edge of the graphene on top of Gate 1. Carriers are injected from the source contact and collimated by the cavity wall, allowing only a small portion of carriers with small incident angle to enter the cavity. (b) When $|n_1| < |n_2|$, the uncollimated carriers that escape the cavity predominantly reach voltage probe B, resulting in a transverse voltage proportional to the current density of uncollimated carriers. At $4|n_1| = |n_2|$, Veselago interference further localizes uncollimated carriers in the cavity. (c) When $|n_1| \geq |n_2|$, the collimation efficiency is not characterized by transverse voltage. (d) Measured transverse voltage as a function of carrier density $n_1$ ($n_2$) in Gate 1 (2). A near-zero transverse voltage is measured at $4|n_1| = |n_2|$ in the midst of high resistance background, as Veselago interference further localizes uncollimated carriers in the cavity. (e) Weak bias dependence is observed for the zero transverse resistance dip, as seen in second-order peaks at $4|n_1| = |n_2|$. (f) As a perpendicular magnetic field is applied beyond 0.3 T, the transverse voltage increases both due to the destruction of Veselago interference as well as the curved trajectory of carriers after passing through the cavity. (g) No significant temperature dependence is recorded from $T = 4$ K ~ 12 K, consistent with previous observations.